\documentclass[aps,prb,twocolumn,showpacs,floatfix,superscriptaddress]{revtex4-1}
\usepackage{graphicx}
\usepackage{multirow}
\usepackage{mathtools}
\usepackage{amssymb}
\usepackage{amsfonts}
\usepackage{mathrsfs}
\usepackage{color}
\usepackage[normalem]{ulem}

\begin{document}
\title{Emerging giant resonant exciton induced by Ta-substitution in anatase TiO$_{2}$: a tunable correlation effect}
\author{Z. Yong}
\affiliation{NUSNNI-NanoCore, National University of Singapore, Singapore 117576}
\affiliation{Singapore Synchrotron Light Source, National University of Singapore, 5 Research Link, Singapore 117603, Singapore}
\affiliation{Department of Physics, National University of Singapore, Singapore 117542}
\affiliation{Department of Electrical and Computer Engineering, National University of Singapore, Singapore 117576}
\author{P. E. Trevisanutto}
\affiliation{Singapore Synchrotron Light Source, National University of Singapore, 5 Research Link, Singapore 117603, Singapore}
\affiliation{Centre for Advanced 2D Materials and Graphene Research Centre, National University of Singapore, Singapore, 117546}
\affiliation{Department of Physics, National University of Singapore, Singapore 117542}
\author{L. Chiodo}
\affiliation{Unit of Nonlinear Physics and Mathematical Modeling, Department of Engineering, Universit\`a Campus Bio-Medico di Roma, Via \'Alvaro del Portillo 21, 00128, Rome, Italy}
\affiliation{Center for Life Nano Science @Sapienza, Istituto Italiano di Tecnologia, Viale Regina Elena 291, 00161, Rome, Italy}
\author{I. Santoso}
\affiliation{NUSNNI-NanoCore, National University of Singapore, Singapore 117576}
\affiliation{Singapore Synchrotron Light Source, National University of Singapore, 5 Research Link, Singapore 117603, Singapore}
\author{A. R. Barman}
\affiliation{Singapore Synchrotron Light Source, National University of Singapore, 5 Research Link, Singapore 117603, Singapore}
\affiliation{Department of Physics, National University of Singapore, Singapore 117542}
\author{T. C. Asmara}
\affiliation{NUSNNI-NanoCore, National University of Singapore, Singapore 117576}
\affiliation{Singapore Synchrotron Light Source, National University of Singapore, 5 Research Link, Singapore 117603, Singapore}
\affiliation{Department of Physics, National University of Singapore, Singapore 117542}
\author{S. Dhar}
\affiliation{NUSNNI-NanoCore, National University of Singapore, Singapore 117576}
\affiliation{Department of Electrical and Computer Engineering, National University of Singapore, Singapore 117576}
\affiliation{Department of physics, School of Natural Sciences, Shiv Nadar University, Gautam Buddha Nagar, P.O. NH-91, Uttar Pradesh 201314, India}

\author{A. Kotlov}
\affiliation{Photon Science at DESY, Notkestra\ss e 85. D-22607 Hamburg, Germany}
\author{A. Terentjevs}
\affiliation{Istituto di Nanoscienze-CNR, Euromediterranean Center for Nanomaterial Modelling and Technology (ECMT), Via per Arnesano 73100 Lecce, Italy}
\author{F. Della Sala}
\affiliation{Istituto di Nanoscienze-CNR, Euromediterranean Center for Nanomaterial Modelling and Technology (ECMT), Via per Arnesano 73100 Lecce, Italy}
\affiliation{Center for Biomolecular Nanotechnologies @UNILE, Istituto Italiano di Tecnologia,Via Barsanti, I-73010 Arnesano, Italy}
\author{V. Olevano}
\affiliation{CNRS, Institut N\'{e}el, Grenoble, France}
\author{M. R\"{u}bhausen}
\affiliation{Institute of Nanostructure and Solid State Physics, Jungiusstrasse 11, 20355, University of Hamburg (Germany).}
\affiliation{Center for Free-Electron Laser Science, Advanced Study Group of the University of Hamburg, Luruper Chaussee 149, 22761 Hamburg \affiliation{NUSNNI-NanoCore, National University of Singapore, Singapore 117576}
(Germany)}
\author{T. Venkatesan}
\email{venky@nus.edu.sg}
\affiliation{NUSNNI-NanoCore, National University of Singapore, Singapore 117576}
\affiliation{Department of Physics, National University of Singapore, Singapore 117542}
\affiliation{Department of Electrical and Computer Engineering, National University of Singapore, Singapore 117576}
\author{A. Rusydi}
\email{phyandri@nus.edu.sg}
\affiliation{NUSNNI-NanoCore, National University of Singapore, Singapore 117576}
\affiliation{Singapore Synchrotron Light Source, National University of Singapore, 5 Research Link, Singapore 117603, Singapore}
\affiliation{Department of Physics, National University of Singapore, Singapore 117542}
\date{\today}
%
%
\begin{abstract}
Titanium dioxide (TiO$_2$) has rich physical properties with potential implications in both fundamental physics and new applications. Up-to-date, the main focus of applied research is to tune its optical properties, which is usually done via doping and/or nano-engineering. However, understanding the role of $d$-electrons in materials and possible functionalization of $d$-electron properties are still major challenges. Herewith, within a combination of an innovative experimental technique, high energy optical conductivity, and of the state-of-the-art  {\it ab initio} electronic structure calculations, we report an emerging, novel resonant exciton in the deep ultraviolet region of the optical response. The resonant exciton evolves upon low concentration Ta-substitution in anatase TiO$_{2}$ films. It is surprisingly robust and related to strong electron-electron and electron-hole interactions. The $d$- and $f$- orbitals localization, due to Ta-substitution, plays an unexpected role, activating strong electronic correlations and dominating the optical response under photoexcitation. Our results shed light on a new optical phenomenon in anatase TiO$_{2}$ films and on the possibility of tuning electronic properties by Ta substitution.
\end{abstract}


\maketitle

\section{Introduction}
Doped or defective titanium dioxide (TiO$_{2}$) exhibits rich physical phenomena in electronic transports and optical properties\cite{1,2,3,4}.  TiO$_{2}$ is opaque in the visible sun light whereas it is very efficient in absorbing ultraviolet (UV) light rendering it interesting especially for photocatalysis applications \cite{5}. The first step of photoexcitation is the formation of electron-hole pair quasiparticles (excitons), which may either recombine or decay into free charges. Eventually  the free charges react with molecules on the surface enhancing photocatalytic effects and formation of reactive free radicals \cite{6}. Excitons, and their spatial behaviour, play therefore a key role for both fundamental physics as well as for applications, but the precise nature and behaviour of excitons in TiO$_{2}$ based materials remains unclear in some respects.

The many-body electron-electron (e-e) and electron-hole (e-h) interactions determine the physical properties of excitons, with different contributions depending on the system and on the considered energy range. Excitons usually occur below direct bandgap in semiconductor and insulator materials, but they may involve higher energy bands in the case of strong electronic correlation. With the recent development of supercomputing and  {\it ab initio} calculations \cite{7,8,9}, theoretical studies have shown that when both e-e and e-h interactions are strongly coupled, they yield to a new type of optical phenomenon, the so-called high-energy resonant excitonic effect. In fact, resonant excitons have been predicted \cite{10, 11} and later observed \cite{12, 13} in two-dimensional graphene. Unlike excitons in conventional semiconductors, the resonant excitons can occur at energies even well above the corresponding optical band gap of the material, and they can be probed directly using high-energy optical conductivity \cite{13}.
A detailed understanding of the role of e-e and e-h interactions in TiO$_{2}$ based materials remains elusive, and resonant excitons have not been observed in the material, mainly because both experimental and theoretical studies at high-energy optical conductivity are challenging and limited in number.

We report in this paper on optical studies of TiO$_2$ doped at different concentrations of Tantalum, via optical conductivity measurements and  {\it ab initio} Time Dependent Density Functional Theory (TDDFT) calculations.
We have observed  resonant excitonic effects in the deep-ultraviolet (DUV) in anatase Ta$_{x}$ Ti$_{1-x}$O$_{2}$ films, with only a small amount of Ta-substitution. A series of unusual phenomena arise, in particular  the spectral-weight transfer from high towards low  energies, and the emergence of an intense resonant exciton at $\sim6$ eV. Based on our theoretical calculations, we relate these effects to a peculiar manifestation of strong e-e and e-h interactions.
The paper is organized as follows: in section \ref{MatMet}, experimental and theoretical-computational used techniques are described. In section \ref{secIII}, optical spectra, both measured and calculated, are described. In section
\ref{Conclusions}, the main conclusions are drawn.

\section{Materials and Methods}\label{MatMet}
Details of samples preparation and characterization, optical conductivity measurements and theoretical calculations are described in this section.
\subsection{Experimental  techniques}
The optical conductivity was obtained using a combination of spectroscopic ellipsometry ($0.5-5.6$ eV) and UV-VUV reflectivity ($3.7-35$ eV) measurements \cite{SI_exp1,SI_exp2}. The spectroscopic ellipsometry measurements were performed in the spectral range between 0.5 and 5.6 eV by using an SE 850 ellipsometer at room temperature.
Three different incidence angles of $60^{\circ}$, $70^{\circ}$ and $80^{\circ}$ from the sample normal were used, and the incident light was $45^{\circ}$ linearly polarized from the plane of incidence. For reflectivity measurements in the high-energy range between 3.7 and 35 eV, we used the SUPERLUMI beamline at the DORIS storage ring of HASYLAB (DESY). The incoming photon was incident at an angle of $17.5^{\circ}$ from the sample normal with linear polarization parallel to the sample surface. The sample chamber was outfitted with a gold mesh to measure the incident photon flux after the slit of the monochromator. The measurements were performed in ultra-high vacuum environment (chamber pressure of $5 \times 10^{-10}$ mbar) at room temperature. Before these measurements, the samples were heated up to $400$ K in ultra-high vacuum to ensure that there were no additional adsorbate layers on the surface of the samples. The obtained UV-VUV reflectivity data were calibrated by comparing it with the luminescence yield of sodium salicylate (NaC$_{7}$H$_{5}$O$_{3}$) and the gold mesh current. These as-measured UV-VUV reflectivity data were further normalized by using the self-normalized reflectivity extracted from spectroscopic ellipsometry \cite{SI_exp3}.
\subsection{Experimental samples and preparations}
Ta$_2$O$_5$ and TiO$_2$ powders with high-purity ($99.999\%$) were ground for several hours before sintering in a furnace at 1000$^{\circ}$C in air for 20 h. Subsequently, target pellets were made and sintered at 1100$^{\circ}$C in air for 24 h. Anatase Ta$_x$Ti$_{1-x}$O$_{2}$ epitaxial thin films (with x=0, 0.018 and 0.038) of thickness 280 nm were deposited on high-quality (001) LaAlO$_3$ substrates by Pulsed Laser Deposition, using a 248 nm Lambda Physik excimer laser with an energy density of 1.8 J cm$^{-2}$ and a repetition rate of 2–10 Hz. Depositions were performed for 0.5–1 h in a stable oxygen partial pressure of $1\times10^{-5}$ Torr while the substrate temperature was maintained at 750 $^{\circ}$C. The chemical and structural properties of the samples were studied by X-ray Photoelectron Spectroscopy, electrical transport measurements, Rutherford backscattering spectrometry (RBS)/channelling, X-ray diffraction (XRD) and time-of-flight secondary-ion mass spectrometry (TOF-SIMS) as reported elsewhere \cite{SI_exp4,SI_exp5}. Ion channelling measurements indicated near perfect substitutional Ta atoms in Ti sites.
\subsection{Theoretical and computational methods}
All ground state electronic calculations are carried out by using Density Functional Theory (DFT) based on Quantum Espresso\cite{espresso} and Abinit \cite{Abinit} codes, with the Perdew-Burke-Ernzerhof (PBE)-GGA approximation for the exchange-correlation functional\cite{PBE}.  Norm conserving pseudopotentials in Troullier-Martins scheme \cite{T-Mpseudo} are used, and semicore electrons are included in Ti and Ta pseudopotentials. The cutoff energy for the expansion of plane-wave basis is up to 170 Ry \cite {17}.
For pristine anatase TiO$_2$, we used a 12$\times$12$\times$8 Monkhorst-Pack k-point mesh sampling the Brillouin zone. For anatase Ta-TiO$_2$ bulk we used supercells with 48 atoms and a 4$\times$4$\times$4 Monkhorst-Pack k-point mesh grid. One Ti atom is replaced with one Ta atom (substitutional doping, modelling a  6.5 $\%$ Ta-doping in TiO$_2$  bulk almost equivalent to the experimental doped sample (3.8 $\%$)).

The excited state calculations have been performed within two approaches: solving the Bethe Salpeter Equation (BSE, which implicitly includes both e-h and e-e interactions), and applying the jellium with gap model (JGM) kernel \cite{28} within the Time Dependent Density Functional Theory (TDDFT). The latter method includes the e-e and e-h interactions maintaining the computational feasibility for such compelling calculations. The complex dielectric function has also been evaluated at the random phase approximation (RPA) level, with electrons and holes treated as independent particles, without correlation. The Bethe-Salpeter Equation has been solved using Yambo code \cite{Marini} and EXC \cite{EXC}. The screening dieletric matrix has been evaluated by using the static inverse dielectric function, with cutoffs of 21 Ry for the correlation (exchange) part, and unoccupied states are summed over 176 empty states. In BSE calculations defined hereafter as Low Resolution (LR), 28 occupied bands and 52 empty bands are included in the diagonalization, to describe the region above 5 eV, on a k-point grid of $4\times$4$\times$2. For the High Resolution (HR) BSE calculation, used to describe in more detail the adsorption threshold, 8 occupied and 8 empty bands are included in the diagonalization, on a 12$\times$12$\times$12 k-point grid. Haydock recursive approach for diagonalization is used, with threshold accuracy of -0.02. The DP-EXC code \cite{dp-code} is used for the TDDFT calculations. In pristine TiO$_2$ bulk, 200 bands are included for the RPA and JGM-TDDFT calculations.

\section{Results and Discussion}\label{secIII}
\begin{figure}
	\includegraphics[width=\columnwidth]{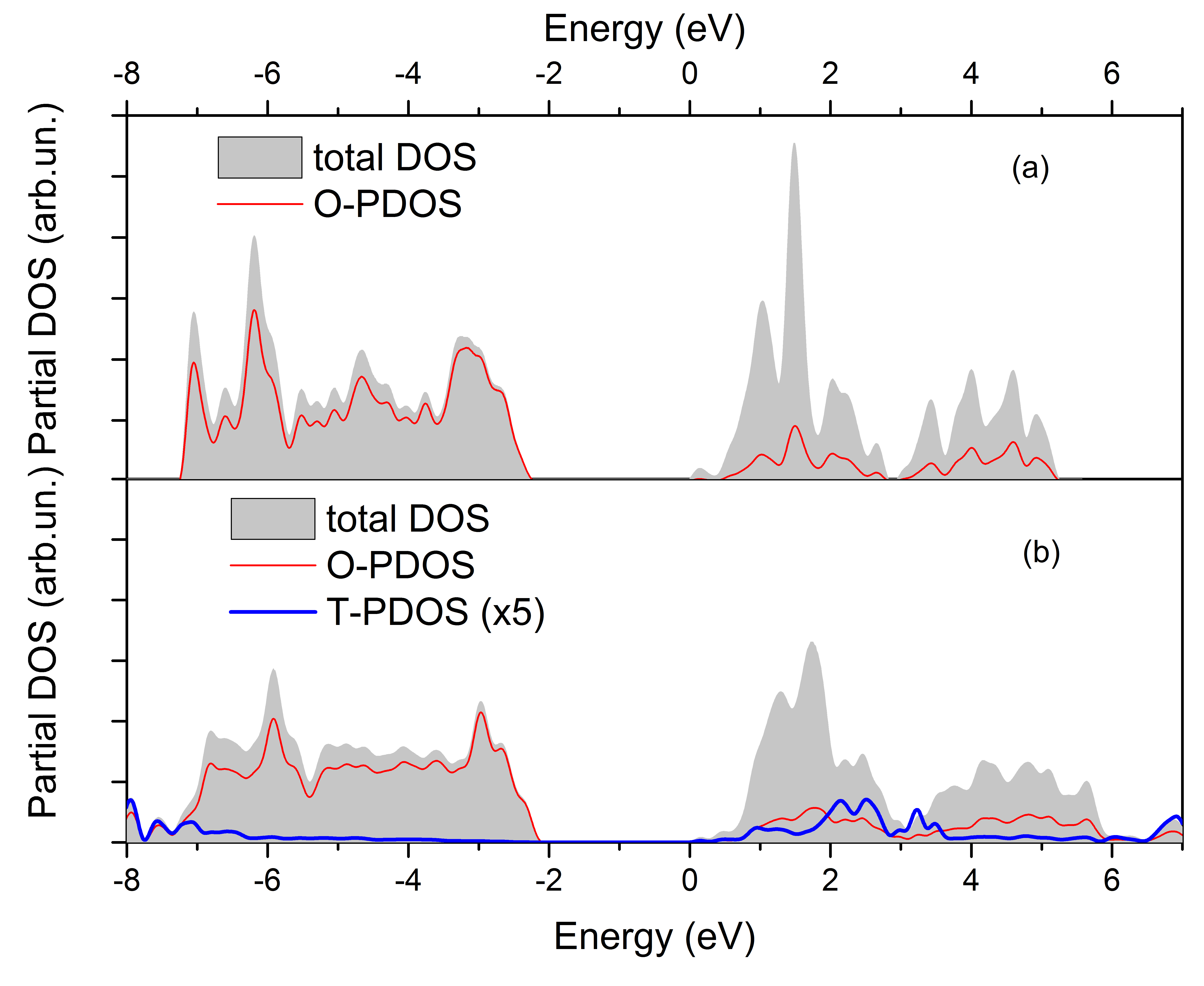}
	\caption{(a) Total DOS of anatase TiO$_{2}$ from DFT calculations. Black line is referred to the total DOS. Red line is the Oxygen partial DOS. (b) DOS of 6.5 $\%$ Ta-substituted anatase TiO$_{2}$. The blue line is the partial DOS of Ta atom multiplied by a factor 5 to make more clear the peaks position.}
	\label{Fig1}
\end{figure}

Here, we provide some general information on pure TiO$_2$, useful in the following discussion on optical conductivity. The electronic ground state structure of pristine and doped TiO$_2$, based on DFT results for total Density of States (DOS) and Partial DOS (PDOS), is shown in Fig.~\ref{Fig1}. The DFT-PBE band gap of pristine TiO$_2$ is $\sim$2.20 eV. The valence band (Fig.~\ref{Fig1}-(a)) mainly consists of O 2p orbitals slightly hybridized with Ti 3d orbitals. The conduction band is comprised by Ti 3d orbitals with a small hybridized amount O 2p orbitals.

\begin{figure}
	\includegraphics[width=\columnwidth]{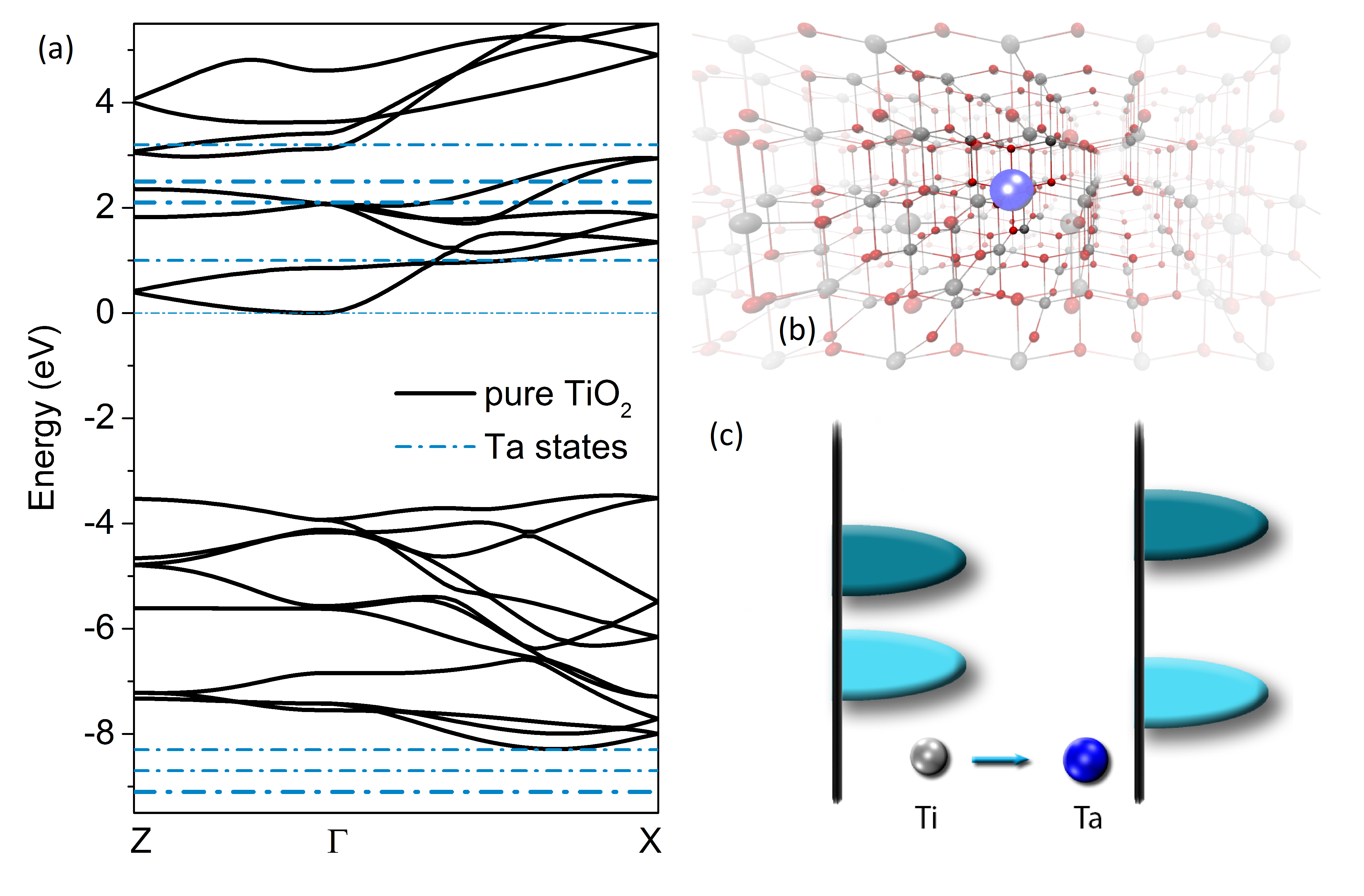}
 	\caption{(a) Anatase band structure (black line), along the high-symmetry directions Z-$\Gamma$-X of Brillouin Zone. A scissor of 1.2 eV has been applied on top of PBE-DFT band structure. Ta-derived states, extrapolated from the PDOS analysis, has been superimposed on the pristine TiO$_2$ band structure (blue dashed line). We can rule out a direct effect of Ta-doping on optical properties, as optical transitions involving Ta-states neither coincide in energy with peaks observed in optical conductivity, neither the PDOS associated to the low Ta-doping here considered could generate the intense optical features we observed. (b) A graphical representation of the long-range correlation effects of the low Ta-doping in TiO$_{2}$ anatase crystal. (c) A simplified cartoon representation of the effects of Ti - Ta substitution that turns on the on-site Coulomb repulsion involving d-Ti and p-O states. The on-site repulsive interaction induces a bandgap opening,  giving optical transitions above the optical gap. }
 	\label{Fig2}
 \end{figure}

In Ta-TiO$_2$ (Fig.~\ref{Fig1}-(b)), the 3d Ta orbitals fill the bottom of conduction band, making the system metallic, and they are hybridized with the adjacent O 2p orbitals up to 8 eV in the conduction band. The proper inclusion of correlation removes this spurious metallicity described by DFT, as shown for Nb-doped rutile \cite{Nb-rutile}.
Main features of electronic band structure (Fig. \ref{Fig2}-(a) ) and optical absorption spectrum of anatase TiO$_2$ have been studied for a long time, and they have been deeply revised and reanalized recently \cite{baldini}.

\begin{figure}
 \includegraphics[width=\columnwidth]{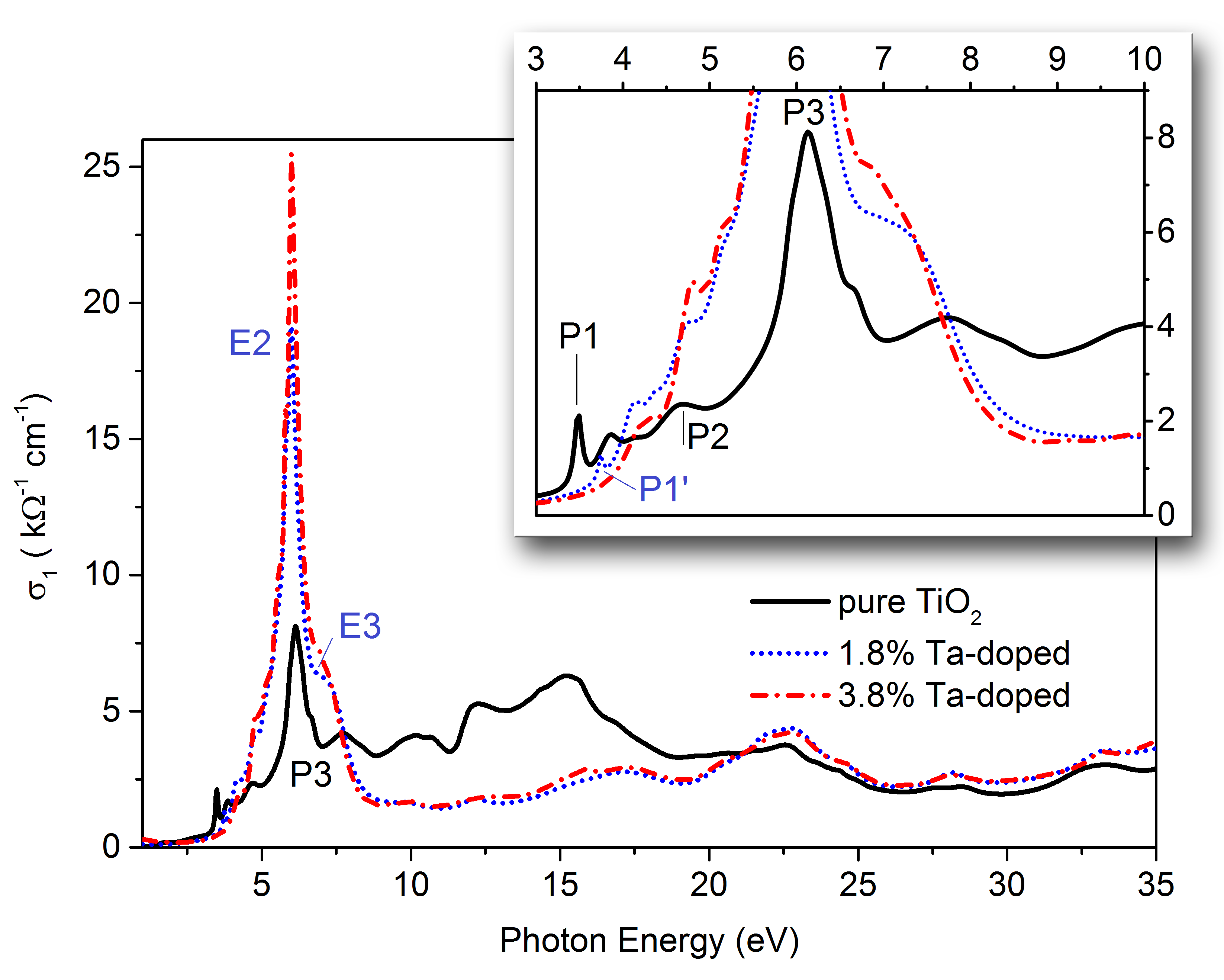}
 \caption{Room temperature measurements of the real part of the optical conductivity for pure TiO$_{2}$ (solid black line), 1.8$\%$ (blue short dotted line), and 3.8$\%$ Ta-substituted TiO$_{2}$ (red dashed-dotted line). The polarization vector is perpendicular to the [001] direction. Inset: details of the real part of the optical conductivity in the VIS and low UV regions.}
 \label{Fig3}
 \end{figure}

Next, we focus on the large spectral changes induced by Ta-doping on the optical response , even for  small amounts of Ta-substitution.
Fig. \ref{Fig3} shows the optical conductivity of TiO$_2$ and Ta-doped TiO$_{2}$ films, at increasing Ta-doping and on a broad energy range, up to $\sim$35 eV. Ta$_{x}$ Ti$_{1-x}$O$_{2}$ films are measured for x = 0, 0.018 and 0.038. For x = 0, the pure TiO$_{2}$ sample, we observe a first sharp optical excitation at 3.48 eV (P1, see inset Fig.~\ref{Fig3}), followed by bulk resonances from 3.85~eV to 4.6~eV (P2) \cite{baldini}. A well-defined large peak, at $\sim$6.12~eV (P3), is a newly observed intense bulk resonance in pure TiO$_{2}$. It is followed by broad and multiple structures up to $\sim$35~eV.

Upon Ta-substitution (x= 0.018) we observe an emerging new giant peak at ~6.0 eV (E2, Fig. \ref{Fig3}), three times more intense than the P3 peak of pure anatase. At ~6.8 eV there is an intense shoulder (E3) of the giant peak E2. The first optical excitation is also affected by Ta-doping, as it occurs at higher energy (~3.75 eV, P1$'$) and reduced in intensity with respect to the pure sample. Furthermore, upon substitutional doping, the spectrum shows a significant reduction of the spectral-weight in a broad energy range ( from $\sim$8~eV to $\sim$20~eV) and a slight spectral-weight gain, singular at even higher energy (from  $\sim$20~eV to $\sim$35.0~eV). For higher Ta-concentration (x=0.038), the E2 at ~6.0 eV peak shows further enhanced intensity, without any significant change in the remaining structures with respect to lower Ta-concentration. To summarize, we have, upon Ta-doping: (i) an anomalous spectral-weight transfer from energies as high as 35 eV towards the 6 eV region; (ii) the emergence of a novel resonant exciton E2 at ~6 eV; and (iii) the strong modification of TiO$_{2}$ optical conductivity, with an augmented optical bandgap.
The optical conductivity here measured in such a broad energy range results crucial to investigate the nature of E2. Based on the optical $f$-sum rule, we find that the total spectral-weight (up to 35 eV) is nearly conserved for all three investigated doping ratios (Fig. \ref{Fig3}). This directly implies that the oscillator strength at ~6.0 eV is coming from spectral-weight transfers of the higher bands, i.e. from ~8 to ~20 eV. Such a collective spectral-weight transfer is a fingerprint of strong electronic correlations \cite{14,15,16}. Our theoretical analysis (see below) shows that the E3 peak in Ta$_{x}$ Ti$_{1-x}$O$_{2}$ has similar origin as the P3 peak at 6.12 eV in undoped TiO$_{2}$, while E2 can be associated to an evolution of the bulk resonance P2. Further, based on our theoretical calculations, we could investigate the role of e-e and e-h interactions, clarifying the nature of the giant exciton E2 and confirming that observed optical behaviour is due to a manifestation of strong e-e and e-h interactions.

\begin{figure}
	 	\includegraphics[width=\columnwidth]{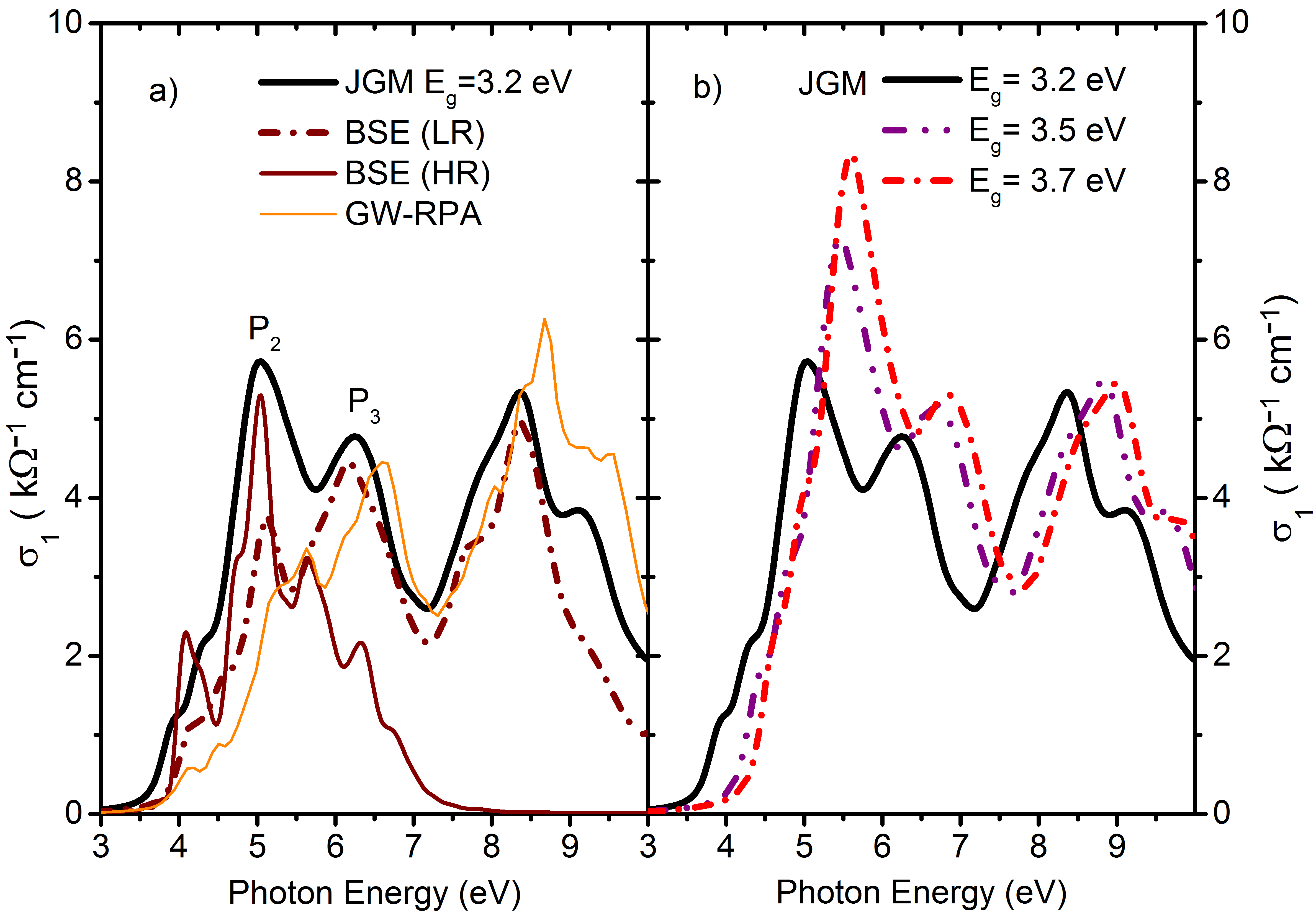}
	\caption{(a) The calculated optical conductivity ([001] in-plane polarization) for anatase TiO$_{2}$ in the GW-RPA approximation (orange solid line), solving the Bethe-Salpeter equation in low resolution (LR-BSE, dark red dashed-dotted line)and High Resolution BSE (HR-BSE, dark red solid line), and within TDDFT using the JGM kernel with E$_g$ = 3.2 eV - experimental indirect bandgap of TiO$_{2}$(black solid line). The HR-BSE calculations include more stringent convergence parameters but with less included conduction bands. The LR-BSE calculations are performed to describe high energy features (beyond 6 eV). Both of them are performed to validate the JGM-TDDFT calculations.
	(b) Optical conductivity ([001] in-plane polarization) for anatase Ta$_{x}$ Ti$_{1-x}$O$_{2}$ calculated by TD-DFT with JGM kernel and E$_g$ of 3.2 eV (solid black line), 3.5 eV (dotted purple line), and 3.7 eV (dashed-dotted red line). Increasing E$_g$ results in an enhanced peak at 6 eV. }
	\label{Fig4}
\end{figure}

In Fig. \ref{Fig4} (a), we show the optical conductivity of the pure TiO$_{2}$ calculated using the three above mentioned theoretical methods. The GW-BSE calculations, taking into account both the e-e and e-h interactions, give rise to bound and resonant excitons or other excitonic effects along with spectral-weight transfers. The comparison between the GW-BSE and GW-RPA results gives therefore a direct measure of the excitonic nature of a resonance. The GW-RPA calculation fails in reproducing the P2 bulk resonance and the structures near the absorption edges, while both GW-BSE and JGM-TDDFT, with some differences in their detail, are able to describe the P1 and P2 peaks. This confirms that the e-e and e-h interactions are significant and important, not only for doped anatase, but even for pure TiO$_{2}$, in agreement with previous results \cite{17,18}. We plot here the optical conductivity, but we note that our GW-BSE result (HR) for optical absorption (not shown) is comparable with previous calculations for TiO$_{2}$ dielectric function \cite{12,13,14,baldini}. From Ref. \cite{17,baldini}, we know that the P1 peak (experimentally at $\sim$3.48 eV) related to bound exciton whereas P2 (at $\sim$4.6 eV) comes from a bulk resonance.

The LR and HR GW-BSE data allow to properly align and identify the JGM-TDDFT spectral features with respect to experimental data, having as reference the P2 peak.
The JGM-TDDFT and HR-GW-BSE coincide in intensity and energy for the peak P2, whereas, for the peak P3 and higher energy features, LR-GW-BSE calculations are in good agreement with the JGM-TDDFT.
Upon Ta-substitution, the solution of the GW-BSE becomes computationally cumbersome. We turn therefore to JGM-TDDFT, which is equally reliable, as just shown in the case of pure TiO$_{2}$, but computationally feasible also for large supercells.

We focus on optical features in the region of the $\sim$6.0 eV, and we use TDDFT to qualitatively study the relationship between E$_{g}$ (and therefore the screening properties of the material), and resonant excitonic effects in Ta$_{x}$ Ti$_{1-x}$O$_{2}$. In Fig. \ref{Fig4}(b), we show JGM-TDDFT results for increasing band-gap values, E$_{g}$ = 3.2, 3.5, 3.7 eV for Ta$_{x}$ Ti$_{1-x}$O$_{2}$. The strong correlation mimicked by E$_{g}$ is reflected in the optical response, in particular in the behaviour of the peak at 6.0 eV. Peak P2 undergoes a redshift of almost 1~eV, and at the same time its intensity increases. Other features in the spectrum (as P3)  undergo a similar shift, but no intensity changes are observed other than for P2.

\begin{figure}
  	\includegraphics[width=\columnwidth]{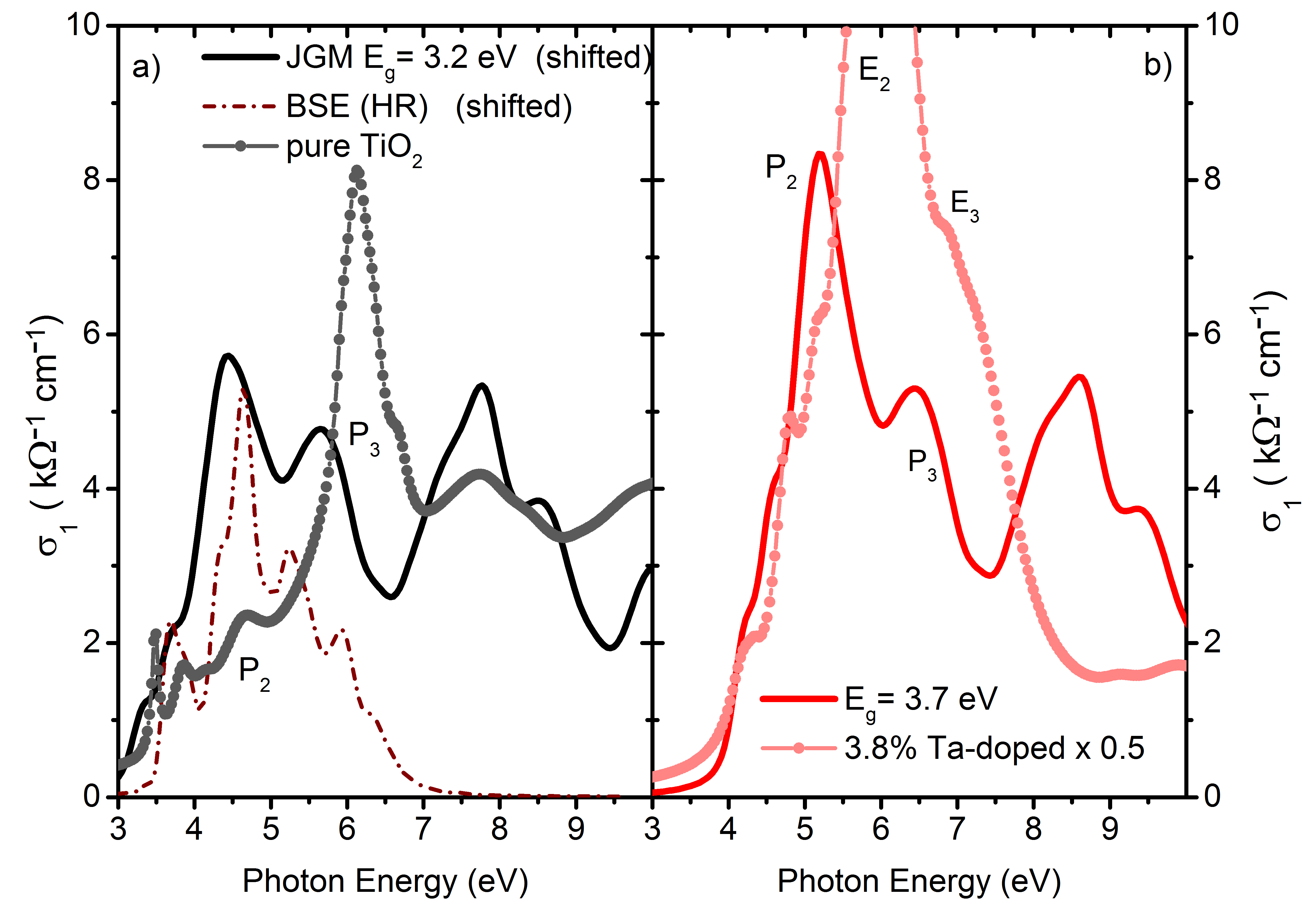}
  		\caption{(a) Comparison between experimental (grey ball lines) and BSE (dot-dashed dark red) and JGM-TDDFT (E$_g$ =3.2 eV, black solid line)   Optical conductivity calculations  for pure anatase TiO$_{2}$. A rigid shift of -0.6 eV has been applied to align the theoretical spectra to experimental data. (b) Experimental optical conductivity (pink balls) and JGM-TDDFT calculations (solid red line) for Ta-substituted ( 3.8 $\%$ TiO$_{2}$ with E$_{g}$=3.7 eV). A rigid shift of -0.6 eV has been applied to align the theoretical spectra to experimental data.}
  	\label{Fig5}
  \end{figure}

In Fig. \ref{Fig5}-(a), we compare the experimental findings with the theoretical calculations. Even tough the results differ in intensity, both P2 and P3 are present. Nevertheless, when the JGM-TDDFT with $E_g=3.7$ eV is compared with the Ta-TiO$_{2}$ optical conductivity (Fig. \ref{Fig5}-b), the theoretical calculations qualitatively suggest that peak P2 is evolving in the E2 exciton at 6.0 eV whereas the P3 peak at $\sim$6.1 eV is transforming in the shoulder E3 at 6.8 eV (Fig \ref{Fig5}(b)). This seems to be counter-intuitive looking at only the experimental optical conductivity results for pure and Ta-substituted TiO$_{2}$, but becomes clear when the proper alignment and assignment of optical features are performed. From the current results, it seems that E2, evolving from P2, is indeed a resonant exciton emerging from an electron-hole continuum which exists at higher energy bands, well beyond a continuum spectrum.

Our JGM-TDDFT calculations display an interplay between E$_{g}$ and the resonant excitons, i.e. larger E$_{g}$ reflects in an enhancement of the resonant excitonic effects. Furthermore, the JGM-TDDFT calculations support the following scenario: the resonant exciton E2 at $\sim$6 eV in experimental spectra can be related to a modification of the electronic structure under Ta-substitution, leading also to the opening of the bandgap. This result is in contrast to a conventional picture, where Ta-substitution would lead to a simple electron doping and metallization of TiO$_{2}$. In fact, our findings imply that Ta-substitution in Ta$_{x}$ Ti$_{1-x}$O$_{2}$ does not act as a conventional dopant, but plays instead an unusual role in enhancing strong electronic correlations. A behavior partaking some similarities with these results has been recently reported for a magnetic-doped TiO$_{2}$ system. Theoretical investigation of magnetic Cr-doped TiO$_{2}$ \cite{19} shows that upon Cr-doping the electronic properties undergo a transformation, and the (initially Charge Transfer insulator) system becomes a strongly correlated Mott-Hubbard crystal. We observe here an optical response which is consequent to an analogous effect in the electronic structure: the excitonic strength resulting in an enhancement of the absorption peak is occurred by the band gap opening. Upon electron doping via Ta-substitution, a possible scenario about the increasing electronic correlation is related to the Ti d-d and Ti d - O p orbital repulsions, but a more detailed analysis is left to frameworks with a better treatment of strongly correlated interactions.
The scenario for Ta-doped TiO$_{2}$ optics presents conceptual similarities to strong correlated materials, as cuprates like doped La$_{2-x}$Sr$_{x}$CuO$_{4}$ \cite{20, 21}. In this respect, TiO$_{2}$ is widely considered an intermediate oxide between the Charge Transfer insulator and Mott-Hubbard regimes \cite{Zaanen_Sawa,Bocquet}. We observe some analogies between our Ta-doped semiconductor spectra, and the optical behaviour of undoped Mott insulators (as cuprates), where intense optical absorption in the DUV are due to transitions from the lower to the upper Hubbard band. In the case of Mott-insulators, the change of E$_{g}$ (or the Mott-gap \cite{18}) as function of doping gives a signature of the e-e correlation. By using the Dynamical Mean Field Theory (DMFT), it has been shown that different percentages of doping enable a change of phase.  The increase of d-states modifies their electronic Density Of States (DOS), and the pseudo-gap material becomes insulating. This results in a strong enhancement of peaks intensity in the optical conductivity.  A similar behavior may be revealed in the present case where the inclusion of Ta d- and f- orbitals seems to have a role in changing the TiO$_{2}$ physics increasing both e-e and e-h correlations.

\section{Conclusions}\label{Conclusions}

In conclusion, we have presented the emergence of an intense resonant exciton induced by Ta-substitution in anatase TiO$_{2}$.
This result is of primarily importance for possible industrial applications.
We argue that in these experimental findings tunable e-e and e-h correlations play a key role in the observed resonant excitons in Ta$_{x}$ Ti$_{1-x}$O$_{2}$ system, and can be used in a model for resonant excitonic effects. Further works will be devoted to improve our qualitative description in a more quantitative agreement with the experimental results.

 It is then also important for future theoretical study to explicitly include the on-site Coulomb repulsion in the optical spectra calculations.



%
\section{Acknowledgments}
This work is supported by Singapore National Research
Foundation under its Competitive Research Funding (NRF-CRP 8-2011-06), MOE-AcRF Tier-2 (MOE2015-T2-1-099), and FRC. We acknowledge the CSE-NUS computing centre, Centre for Advanced 2D Materials and Graphene Research Centre for providing facilities for our numerical calculations. We also acknowledge the National Research Foundation, Prime Minister’s Office, Singapore, under its Medium Sized Centre Programme and Competitive Research Funding (R-144-000-295-281). L.C. acknowledges M. Lauricella and J. Sofo for useful discussions. Authors acknowledge F. Da Pieve for a critical reading of the manuscript. Z.Y., P.E.T. and L.C. contributed equally.

\end{document}